\documentclass{svjour3}                     
\smartqed  
\usepackage{graphicx}
 \journalname{Bulletin of Mathematical Biology}
\begin{document}

\title{Electromigration dispersion in Capillary Electrophoresis
\thanks{supported by the NIH under grant  R01EB007596.}
}


\author{ Zhen Chen    \and  Sandip Ghosal}


\institute{Z. Chen \at
           Northwestern University, Dept. Mech. Eng., 2145 Sheridan Road, Evanston, IL 60208\\
              Tel.: +1 847 491 3663\\
              Fax: +1 847 491 3915\\
              \email{zhenchen2011@u.northwestern.edu}         
           \and
           S. Ghosal \at
              Northwestern University, Dept. Mech. Eng., 2145 Sheridan Road, Evanston, IL 60208\\
              Tel.: +1 847 467 5990\\
              Fax: +1 847 491 3915\\
              \email{s-ghosal@northwestern.edu}           
}

\date{Received: date / Accepted: date}

\maketitle

\begin{abstract}
In a previous paper (S. Ghosal and Z. Chen {\it Bull. Math. Biol.} 2010 {\bf 72},  pg. 2047) it was shown that the evolution of the solute concentration 
in capillary electrophoresis is described by a nonlinear wave equation that reduced to Burger's equation if the nonlinearity was weak.
It was assumed that only strong electrolytes (fully dissociated) were present. In the present paper it is shown that the same governing 
equation also describes the situation where the electrolytic buffer consists of a single weak acid (or base). A simple approximate formula is derived 
for the dimensionless peak variance which is shown to agree well with  published experimental data.
\end{abstract}

\section{Introduction}
Capillary electrophoresis (CE) is a technique for separating a mixture of macro-ions in aqueous 
solution by exploiting the fact that the migration velocity in an applied field depend on the size and 
charge of the molecule. It is a widely used laboratory tool  in  bio-analytical chemistry.
Further background information may be found in the authors' earlier paper in this journal~\cite{ghosal_chen10} (henceforth refereed to as 
GC) and some of the references cited there. There are also several textbooks devoted to the subject~\cite{czebook1,czebook2}. 

The mathematical formulation of the problem consists of a set of coupled equations describing the transport of ions in response 
to electric fields and diffusive fluxes. The electric field in turn is determined by the concentration distribution of ions.
Coupling to hydrodynamics could occur due to electro-osmosis, but is neglected here for simplicity. Electromigration 
dispersion (EMD) is caused by variations in the  local electrical conductivity in the vicinity of the solute peak which 
gives rise to a nonlinearity in the equation for solute concentration. 

Simple one dimensional mathematical models of electromigration dispersion neglecting the effects of diffusion 
have been studied by various authors~\cite{mi_ev_ve_79a,mikkers_ac99,math_th_elph_bk}. The problem is 
reduced to a single nonlinear hyperbolic equation for the concentration of sample ions. Analytical and numerical solutions describe the  characteristic 
wedge shaped profile observed in experiments~\cite{gas09,thormann09}. 
The restriction to zero diffusivity was removed recently by the authors~\cite{ghosal_chen10} 
 who considered a `minimal system' of  three ions -- the sample ion, co-ion and counter-ion -- all being  strong electrolytes (fully dissociated).
The diffusivities of the three ionic species were considered non zero but equal. The sample concentration was then shown to 
obey a one dimensional nonlinear advection diffusion equation which reduced to Burgers' equation if the sample concentration was not too high. Thus, 
the concentration profile could be obtained analytically as a function of position and time, providing useful insights
into the nature of electromigration dispersion.

The solutes of interest in CE are often biological molecules for which the charge is quite sensitive to the pH of the surrounding 
electrolyte. Many molecules are also unstable outside a narrow pH range. Thus, CE must be performed in a medium where 
the pH is kept as constant as possible. In order to maintain a stable pH an electrolytic 
buffer containing  a weak acid or base is used as the background electrolyte. The buffer often contains several 
ionic species as well as other additives to 
achieve different functions (e.g. prevent adsorption to capillary walls). Thus, the simplifying assumption made by GC
that the buffer consists of a single strong electrolyte is often not consistent with laboratory practice. 

In this paper we show that the 
one dimensional model derived by GC for strong electrolytes may also be applied to an idealized model of a buffer 
consisting of a single weak electrolyte. The only change required is in the definition of a parameter in the model that 
characterizes the strength of EMD effects.
 A simple formula for the ``number of theoretical 
plates'' -- a dimensionless measure of peak dispersion --  is derived and is shown to agree well with published 
experimental data.

The rest of the paper is organized as follows: the theory of EMD  in the presence of a weak electrolytic buffer is provided in the next section 
followed by a discussion of the experimental work and comparison with the theory. A summary of our results and discussion of the validity of our underlying 
assumptions is provided in the concluding section.

\section{Theory}
For definiteness, we consider an acidic analyte, $H_{n}A + nH_{2}O \rightarrow  nH_{3}O^{+} + A^{n-}$ in aqueous solution buffered with a weak acid 
$HX + H_{2}O \rightleftharpoons H_{3}O^{+} + X^{-}$.
The auto ionization of water is neglected. The analyte is considered to have a fixed charge ($ze$), but, the buffer could exist 
in either a charged ($X^{-}$) or neutral ($HX$) state. Cationic analytes or buffers with multivalent ions can be accommodated easily in our analysis 
but we only discuss the simplest situation for convenience. Further, we assume that all ions have the same mobility ($u$), and therefore, 
identical diffusivity ($D$), and that the transport problem is entirely one dimensional. In particular, we assume that 
electroosmotic flow, when present, may be 
described as an advection of the ions with a constant velocity $u_{0}$ in the axial direction. We will return to the question of the 
validity of these assumptions in the concluding section.

\subsection{Derivation of a reduced system}

The coupled equations describing the concentrations of hydrogen ions ($c_{+}$), acid ions ($c_{-}$), sample ions ($c$) and the neutral form
$HX$ ($c_{0}$), are then 
\begin{eqnarray} 
\frac{\partial c_{+}}{\partial t} + \frac{\partial}{\partial x} \left[ (u_{0} + euE) c_{+} \right] &=& D \frac{\partial^{2} c_{+}}{\partial x^{2}} + r, \label{T+}\\
\frac{\partial c_{-}}{\partial t} + \frac{\partial}{\partial x} \left[ (u_{0} - euE) c_{-} \right] &=& D \frac{\partial^{2} c_{-}}{\partial x^{2}} + r, \label{T-}\\
\frac{\partial c_{0}}{\partial t} + u_{0} \frac{\partial c_{0} }{\partial x}   &=& D \frac{\partial^{2} c_{0}}{\partial x^{2}} - r
, \label{T0}\\
\frac{\partial c}{\partial t} + \frac{\partial}{\partial x} \left[ (u_{0} + zeuE) c \right] &=& D \frac{\partial^{2} c}{\partial x^{2}}, \label{T}
\end{eqnarray}
where $E$ is the local electric field, $x$ is the distance along the capillary, $t$ is time, $e$ is the electronic charge and 
$r =  k_{d} c_{0} - k_{a} c_{+}c_{-}$ is the net dissociation rate of $HX$. Here, $k_{a},k_{d}$ are 
constants that characterize the rates for the forward and reverse reactions of the weak acid. We will assume that 
the x-axis points in the direction of peak motion from the 
inlet towards the detector. These equations differ from those considered by GC in the presence of the 
source terms on the right hand sides of the first three transport equations and in the existence of the additional variable, $c_{0}$.
 Since characteristic spatial scales are always much larger than 
the Debye length, local electro-neutrality holds. Thus,
\begin{equation} 
c_{+}-c_{-}+zc = 0.
\label{LEN}
\end{equation}
If we multiply equations (\ref{T+})-(\ref{T}) by the respective ionic charges, sum them, and use equation (\ref{LEN}), we get an 
equation that describes the constancy of electric current 
\begin{equation} 
 \frac{\partial}{\partial x} \left[ e^{2}u ( c_{+} + c_{-} + z^{2} c ) E \right] = 0.
 \label{conserve_curr}
 \end{equation} 
Note that the net contribution from the diffusive fluxes as well as those from the source terms vanish exactly. 
Equation (\ref{conserve_curr}) may then be integrated;
\begin{equation} 
 ( c_{+} + c_{-} + z^{2} c ) E = 2 c_{\infty} E_{\infty}, 
  \label{current}
 \end{equation}
 where $E_{\infty}$ and $c_{\infty}$ are the electric field and cation (or anion) concentration respectively far away from the peak.
 
 Equations (\ref{LEN}) and (\ref{current}) provide two algebraic relations among the five dependent variables 
 $c_{+}, c_{-}, c_{0}, c$ and $E$. To reduce the system of equations (\ref{T+}) -- (\ref{T}) and (\ref{conserve_curr}) to 
 a single one dimensional equation we must seek more such algebraic relations. A third relation is provided by an 
 approximation first introduced by Saville and Palusinski~\cite{saville_theory_1986}: the time scale associated with the
dissociation-recombination reactions represented by the last  term on the right hand side of equation (\ref{T0}) is 
so small compared to all transport time scales that equation (\ref{T0}) may effectively be replaced by 
\begin{equation} 
 r =  k_{d} c_{0} - k_{a} c_{+}c_{-}= 0 
\label{chem_eq}
\end{equation}
or $k c_{0} = c_{+}c_{-}$, where $k = k_{d} / k_{a}$ is the acid dissociation constant. Thus, locally, the acid is in equilibrium 
with its dissociation products. 

In GC it was shown that the Kohlrausch function $K = (c_{+} + c_{-} + c)/u$ is a passive scalar that spreads only by diffusion 
(or, if electroosmosis is present, is also advected with a constant velocity $u_{0}$). Therefore, electrophoretic migration relative to the fluid rapidly
advects the solute peak into a region where $K$ is essentially equal to its far field unperturbed value. This provided the additional 
algebraic relation (the constancy of $K$) that finally enabled the reduction of the transport equations to a one dimensional system. This approach 
does not work in the current problem, as it may be readily shown from the transport equations (\ref{T+}), (\ref{T-}) and (\ref{T}),
that the evolution equation for $K$,
\begin{equation} 
\frac{\partial K}{\partial t} + u_{0} \frac{\partial K}{\partial x} = D \frac{\partial^{2} K}{\partial x^{2}} + \frac{2 r}{u},
\end{equation} 
now has a source term, and $K$  is therefore no longer a passive scalar.

Fortunately  however, a fourth algebraic relation is obtained if one assumes that the buffering action 
of the weak acid $HX$ is ``perfect'' so that 
\begin{equation} 
c_{+} = c_{\infty}.
\label{Hconst}
\end{equation} 
The justification of the approximation (\ref{Hconst}) is rooted in the theory~\cite{helfferich_ion_1995} of acid-base equilibria where 
it is shown that in a mixture of a strong ($HA$) and weak ($HX$) acid,
 the perturbation  in the hydrogen ion concentration is small when the buffering capacity of the background electrolyte is large. 
 The buffering capacity is a maximum, if the solution pH (-$\log c_{+}$) is equal to the pK$_{a}$ (-$\log k$) of 
 the acid and drops sharply if the pH differs by more than one unit from the pK$_{a}$.  
 In practice, the buffer consists of a mixture of the weak 
 acid and its salt so that the pH and  buffer concentration can be independently controlled. At a given pH, the buffering capacity 
 increases with buffer concentration. 
 The buffer concentration cannot be made too large however, as this increases 
 the solution conductivity thereby reducing separation efficiency. 
 In the context of our model represented by  equations (\ref{T+}) - (\ref{T}) and (\ref{conserve_curr}), the significance of the 
 assumption (\ref{Hconst}) becomes clear if we use equations (\ref{LEN}) and (\ref{chem_eq}) to express $c_{+}$ in terms 
 of the concentrations $c$ and $c_{0}$:
 \begin{equation} 
 c_{+} = \sqrt{kc_{0} + z^{2} c^{2} /4} - zc/2.
 \label{eq4c+}
 \end{equation}
 Equation~(\ref{eq4c+}) may be replaced with equation~(\ref{Hconst}), if 
 \begin{equation} 
 kc_{0} \gg z^{2} c^{2},
 \label{large_buffer}
 \end{equation} 
that is, if the amount of undissociated acid in the sample zone is high 
enough to act as a ``buffer'' against pH variations.

 With four algebraic relations among the five dependent variables, equations (\ref{T+}) -- (\ref{T}) and (\ref{conserve_curr}) 
 may now be reduced to a single equation for the normalized solute concentration $\phi = c / c_{\infty}$\,:
 \begin{equation} 
 \frac{\partial \phi}{\partial t} + \frac{\partial}{\partial x} \left[  \left( u_{0} + \frac{v}{1 - \alpha \phi} \right) \phi \right] = 
 D \frac{\partial^{2} \phi}{\partial x^{2}},
 \label{evolvephi}
 \end{equation} 
 where $v=zeuE_{\infty}$ is the electromigration velocity of an isolated solute ion and 
 \begin{equation}
 \alpha = - \frac{1}{2} z(1+z).
 \label{newalpha}
 \end{equation}
Equation (\ref{evolvephi}) is identical 
 to the corresponding equation for strong electrolytes (equation (13) in GC; except there 
 $u_0 = 0$). The only difference is in the definition of $\alpha$ which is given by equation~(\ref{newalpha})
instead of equation~(14) of GC, which is  $\alpha = (1-z^{2})/2$ for a univalent background electrolyte.
 
\begin{figure}[t]
\includegraphics[width=0.8\textwidth]{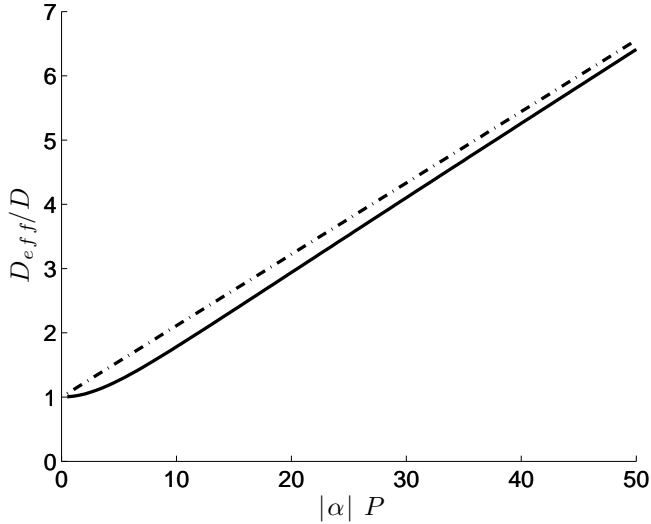}
  \caption{Comparison of the normalized effective diffusivity given by equation~(\ref{Deff}) [solid line] with that 
  predicted by the approximate form equation~(\ref{Deff1}) [broken line].}
  \label{Dratio}
\end{figure}
\subsection{Dispersion}
Equation~(\ref{evolvephi}) is analytically solvable in certain limits and its consequences were
discussed at length in GC. Here we are interested in the peak dispersion of the solute as 
measured by $N = L_{d}^{2} / \sigma^{2}$ where $L_{d}$ is the distance between the injection point and the 
detector, and, $\sigma^{2}$ is the 
variance of the concentration when the peak reaches the detector. One of the consequences of equation (\ref{evolvephi}) is that if 
$\phi$ is  not too large, the variance of the peak increases 
in proportion to the time so that the spreading may be described in terms of an effective diffusivity 
\begin{equation} 
D_{\mbox{eff}} = 2 D \left( \frac{F_{2}}{F_{0}} - \frac{F_{1}^{2}}{F_{0}^{2}} \right) 
\label{Deff}
\end{equation} 
where $F_{n}$ is the $n$ th moment of a certain function $F(x)$ defined by equation (23) in GC. 
The degree of sample loading is characterized by a length $\Gamma = \int_{-\infty}^{+\infty} \phi(x,t) \, dx$
and the relative importance of diffusion is characterized by a Peclet number $P = \Gamma |v| / D$. 
The effective diffusivity has the following asymptotic forms at small and large values of the Peclet 
number $P$\,:
\begin{equation}
D_{\mbox{eff}} \sim \left\{ 
\begin{array}{ll}
D  & \mbox{if  $P \ll 1$},\\
\frac{1}{9} |\alpha v| \Gamma   & \mbox{if $P \gg 1$}.
\end{array}
\right.
\label{Deffasymp}
\end{equation} 
 For the purpose of practical applications, equation (\ref{Deff}), which requires the numerical 
evaluation of integrals may be replaced by the simpler formula 
\begin{equation} 
D_{\mbox{eff}} = D + \frac{1}{9} |\alpha v| \Gamma.  
\label{Deff1}
\end{equation} 
The ratio $D_{\mbox{eff}}/D$ given by equation~(\ref{Deff}) depends solely on $|\alpha| P$.
The approximation to $D_{\mbox{eff}}$ given by equation (\ref{Deff1}) approaches the theoretical 
value given by equation~(\ref{Deff})  in the asymptotic limits of
small as well as large Peclet numbers. In Figure~1 we plot $D_{\mbox{eff}}/D$ as a function of $|\alpha| P$ evaluated 
using equation~(\ref{Deff}) as well as the approximate form equation~(\ref{Deff1}). It is seen that the approximate 
form provides an excellent approximation for all values of $|\alpha| P$ while having the advantage 
of algebraic simplicity.

The number of theoretical plates as a function of the magnitude of the applied voltage, $V$, may be calculated from 
 the classical result~\cite{czebook1} 
 \begin{equation}
 N = \frac{(\mu_{0} + \mu) V}{2 D},
 \label{JLeq}
 \end{equation}
where $\mu_{0}$ is the electroosmotic mobility and  $\mu = zeu$ is the electrophoretic mobility of the sample ion.
Equation~(\ref{JLeq}) may be generalized by replacing the 
diffusivity, $D$, by the effective diffusivity $D_{\mbox{eff}}$. We also allow for an initial peak variance 
($\sigma_{0}^{2}$), and, an inlet to  detector distance ($fL$) that is slightly smaller than the capillary length ($L$).
Thus,
\begin{equation} 
N = \frac{f A V}{1 + BV} 
\label{N}
\end{equation} 
where the constants $A$ and $B$ are respectively 
\begin{equation} 
A = \frac{(\mu_{0} + \mu)}{2 D} 
\label{A}
\end{equation} 
and 
\begin{equation} 
\frac{B}{A} = \frac{\sigma_{0}^{2}}{f^{2} L^{2}} + \frac{2}{9} | \alpha |  \frac{| \mu |}{\mu_{0} + \mu} \frac{\Gamma}{fL}.
\label{B}
\end{equation} 
We will now compare equation (\ref{N}) for $N$ 
with published experimental data.

\begin{figure}[t]
\includegraphics[width=0.8\textwidth]{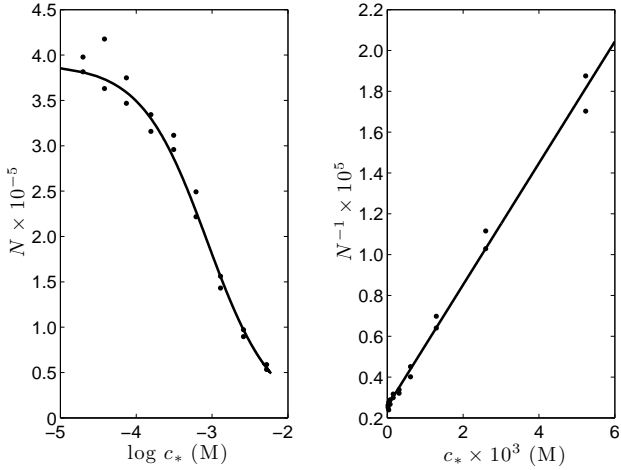}
  \caption{Symbols are data from Figure~3 of Lukacs and Jorgenson~\cite{lukacs_capillary_1985} re-plotted as $N^{-1}$ vs. $c_{*}$ on 
  the right panel. The solid line is the best fit linear regression $N^{-1} = a + b c_{*}$.}
  \label{fitdata}
\end{figure}
 \section{Experiments}
 Jorgenson and Lukacs~\cite{jorgenson_zone_1981} derived equation~(\ref{JLeq}) 
 as a simple model of  peak dispersion. The experimental data that they presented however, showed the expected linear 
 dependence of $N$ on $V$, but only over a limited range of $V$. At large $V$, the linear dependence was found to exhibit 
 a saturation effect so that $N$ approached a limiting value rather than increase indefinitely. Jorgenson and Lukacs 
 attributed this to the effect of Joule heating in the capillary. Delinger {\it et al.}~\cite{delinger_influence_1992}
 later re-interpreted Jorgenson and Lukacs data and suggested that 
 the saturation was more likely due to the initial variance ($\sigma_{0}$) of the injected zone. They also 
 presented additional experimental data to support their interpretation. 
 Equation~(\ref{B}) shows that electromigration dispersion is another possible reason for the observed saturation 
 of $N$ at large $V$. Whether the initial variance or the electromigration effects dominate will depend on the 
 experimental parameters. As the voltage is increased further, Joule heating results in a decrease of $N$.
  In some experiments~\cite{issaq_effect_1991-1}
 Joule heating masks any effect of electromigration and no distinct plateau region is observable in the $N$ vs. $V$ curve.
 Instead, the $N$ vs. $V$ curve displays a single maximum.
Thus, in any comparison of equation~(\ref{N}) with experimental data, care must be taken to distinguish electromigration 
effects from those due to the initial variance and Joule heat. 
Equation~(\ref{B}) suggests that this can be done in the clearest manner if one 
varies the sample loading $\Gamma$ while keeping all other experimental parameters invariant. In this 
situation a linear dependence of $N^{-1}$ with $\Gamma$ should be observed. Such an experiment was published 
by Lukacs and Jorgenson~\cite{lukacs_capillary_1985} and we shall compare equation~(\ref{N}) with their data.

Specifically, we use the data shown in Figure~3 of the paper by Lukacs and Jorgenson~\cite{lukacs_capillary_1985} (henceforth 
referred to as LJ) where the solute 
was Dansyl-isoleucine in a 0.05 M phosphate buffer at pH 6.86.
In the experimental setup, the solute concentration ($c_{*}$)  alone was varied. In each run, 
the same length of plug ($\ell$) was injected into the capillary by electrokinetic injection. Thus, the initial 
profile $\phi(x,0)$ was a square wave, and therefore, $\Gamma = c_{*} \ell / c_{\infty}$ and $\sigma_{0}^{2} = \ell^{2} / 12$.
 According to equation~(\ref{N}),   $N^{-1}$ should have a linear dependence on the concentration: $N^{-1} = a + b c_{*}$ where 
\begin{eqnarray} 
a &=&  \frac{2 D}{fV(\mu_{0} + \mu)} + \frac{1}{12  f^{2}} \frac{\ell^{2}}{ L^{2}}  \label{fita}\\
b &=& \frac{2}{9}  | \alpha | \frac{| \mu |}{ \mu_{0} + \mu } \frac{\ell}{fL} \; \frac{1}{c_{\infty}}
\label{fitb}
\end{eqnarray}
The symbols in Figure~\ref{fitdata} are the experimental data which does show that $N^{-1}$ has a linear dependence 
on $c_{*}$. Fitting a straight line through the data points gives $a = 2.57 \times 10^{-6}$, $b = 2.98 \times 10^{-3} \: M^{-1}$.

The reported values of the various constants appearing in equations~(\ref{fita}) and (\ref{fitb}) are summarized in Table~1.
%
\begin{table}
\caption{Experimental parameters corresponding to the data shown in Figure~\ref{fitdata}}
\label{tab:1}       
\begin{tabular}{cccccccc}
\hline\noalign{\smallskip}
L (cm) & f & V (kV) & $\mu_0~(\mbox{cm}^2/Vs)$ & $\mu~(\mbox{cm}^2/Vs)$ & $D ~(\mbox{cm}^2/s)$ & $z$ & $ c_{\infty} (M)$\\ 
\noalign{\smallskip}\hline\noalign{\smallskip}
100  & 0.80 & 30 & $4.79 \times 10^{-4}$ & $- 1.37 \times 10^{-4}$ &  $5.35 \times 10^{-6}$ &  $- 0.65$ & $0.025$ \\
\noalign{\smallskip}\hline
\end{tabular}
\end{table}
The value of the electroosmotic mobility, $\mu_{0}$ at pH 6.86 for pyrex was read from Figure~4 of LJ. The electrophoretic mobility, $\mu$, diffusivity, $D$, and valence\footnote{Dansyl-isoleucine has a single negative charge on the carboxyl group but its effective charge is likely reduced by shielding effects.} $z$ of Dansyl-isoleucine reported in Table~1 were taken from the paper by Walbroehl and Jorgenson~\cite{walbroehl_capillary_1989}, where these parameters were measured under settings identical to those employed in LJ. The buffer strength of our idealized buffer $HX$ when $pH = pK_{a}$, is, by the 
Henderson-Hasselbalch relation $[HX] + [X^{-}] = 2 c_{\infty}$. Equating this to the reported buffer strength of $0.05$ M gives the value 
of $c_{\infty}$ indicated in Table~1.
Unfortunately, the injection Voltage ($V_{*}$) and injection time ($\tau_{*}$) for the electrokinetic injection process were not reported, 
so that, $\ell$ is not known.
However, if we use the value of $a$ obtained from the best fit line 
in Figure~\ref{fitdata}, we get $\ell = 0.31$ cm. From the relation $\ell = (\mu + \mu_{0}) V_{*} \tau_{*} /L$ we can determine that 
$V_{*} \tau_{*} \approx 90 ~\mbox{kV} \cdot \mbox{s}$, which appears consistent with what is typically reported in 
similar experiments~\cite{delinger_influence_1992}. 
If we substitute this value of the injection plug length, $\ell$, into equation~(\ref{fitb}) and use the value of $b$
found from the linear regression, we deduce that $\alpha = 0.22$. 
The phosphate buffer is more complex than the idealized model of a buffer considered here. However, 
if we put $z=-0.65$ in equation~(\ref{newalpha}) we get $\alpha \approx 0.1$. If the background species was a fully 
dissociated 1-1 electrolyte we would have (GC) $\alpha = (1-z^{2})/2 \approx 0.3$. These numbers are commensurate 
with the value computed from the experimental data using equation~(\ref{fitb}), which suggests that the value of $\alpha$ is probably 
not very sensitive to the simplifications adopted to model the background electrolyte.

\section{Conclusion}
The problem of electromigration dispersion of a solute was considered in the presence of an idealized 1-1
weak electrolytic buffer. The analysis complements earlier work (GC) where the background electrolyte 
was regarded as fully dissociated. It was shown that in both models the solute transport is described by the same 
one dimensional transport equation; they differ only in the definition of the parameter $\alpha$ characterizing the strength of 
EMD effects. A simple expression for the number of theoretical plates was derived by replacing the analytical expression 
for the effective solute diffusivity by an approximate form. This expression for $N$ was then compared with published 
experimental data and good agreement was found. 

The observed agreement between theory and experiment requires further explanation, since the phosphate 
buffer used in the experiment is a complex multi-ion mixture and does not correspond to either the strong 
electrolyte model of GC or the idealized weak electrolyte model considered here. The explanation must be, that, the fact that 
the concentration is governed by equation~(\ref{evolvephi}) is independent of the specific model assumed for the buffer.
This is quite plausible, since to derive equation~(\ref{evolvephi}) all that is required is 
that one must be able to find $N-1$ linear algebraic relations among the $N$ chemical species in solution.
The analysis presented 
in this paper and in GC are simply two alternate ways in which this can be accomplished through different assumptions 
about the nature of the background electrolyte. There are of course other ways to a similar end. For example, in a 
more complicated multi-component background electrolyte, in addition to local electro-neutrality, one 
may assume local equilibrium for a subset of the species and set the concentrations of certain other species to zero or to a constant 
value depending on the details of the buffer chemistry. Alternatively, one could take a more 
heuristic view point and simply say that the local migration speed of a solute ion is a function of the local solute concentration, 
and, if the solute concentration is not too high, this function may be linearized: $v_{e} (\phi) = v [ 1 + \alpha \phi + \cdots ]$. In this approach $\alpha$ would 
be an empirical parameter. Our analysis suggests, that in general, $\alpha$ may be well
approximated by a parabolic function of the solute valence $\alpha = \alpha_{0} + \alpha_{1} z + \alpha_{2} z^{2}$ where 
the coefficients $\alpha_{0},\alpha_{1},\alpha_{2}$ depend on the buffer composition. For a 1-1 strong electrolyte, we found (GC),
$z_0 = - z_{2}=1/2$ and $z_{1}=0$, whereas for a 1-1 weak electrolyte, $z_{1}=z_{2}=-1/2$ and $z_{0}=0$. Determining the 
coefficients $\alpha_{0},\alpha_{1}$ and $\alpha_{2}$ experimentally may be an efficient way of characterizing the dispersive 
properties of electrolytic buffers in the laboratory.

In arriving at equation~(\ref{N}), the local ion migration velocity was linearized: $v_{e}(\phi) = v /(1 - \alpha \phi) \approx v [1 + \alpha \phi]$
by neglecting quadratic and higher powers of $\phi$.  This is a valid approximation if $\phi \ll 1$. 
Furthermore, the smallness of $\phi$ is also inherent in the assumption of a ``perfect buffer'', equation (\ref{Hconst}). Indeed, 
if we require that equations (\ref{Hconst}), (\ref{LEN}) and (\ref{chem_eq}) be consistent with the requirement (\ref{large_buffer}),
then we must have $1 + z \phi \gg z^{2} \phi^{2}$, that is, $\phi$ must be small in comparison 
to unity. In the experiment, even at the highest solute concentrations used, 
$\phi \sim c_{*} / c_{\infty} = 0.006 M / 0.025 M = 0.24$. Thus, $\phi \ll 1$ is a valid 
approximation in the experiments of LJ.

In the current analysis as well as in GC, the diffusivity of all ionic species were presumed equal. 
This is not an entirely unreasonable assumption, since the diffusivities decrease weakly with molecular mass. For 
example, the diffusivity of Dansyl-isoleucine is only about three times larger than that of the sodium ion though 
its molecular mass is about sixteen times as great. The exception is the hydrogen ion which has a diffusivity 
about a factor of ten 
higher
than Dansyl-isoleucine. However, in the phosphate buffer the cations responsible for conducting
current are primarily the sodium ions, the hydrogen ion concentration is actually very small. For macro-ions with molecular 
weights in the range of kilo Daltons as well as for colloidal particles, differential diffusivity could turn out to be important. 
Such effects can be accounted for in our theory 
by including the diffusive current in the equation for current conservation, equation~(\ref{conserve_curr}). This would 
make the electric field dependent not only on the solute concentration but also on its gradient. The resulting one 
dimensional model would include the effects of differential diffusion, but one may lose the convenience of 
having an analytically solvable governing equation.

The pyrex capillary employed in the experiment generated strong electroosmotic flow which was accounted for 
in a simplistic manner as a uniform advection along the capillary. However, the perturbation in the electric field 
generated by the solute peak also perturbs the slip velocity at the capillary wall which results in radial shear 
and consequent Taylor-Aris dispersion by a well known mechanism~\cite{electrophoresis_review_ghosal}.
It may be shown~\cite{ghosal_chen_emd_eof} that this effect can be accounted for by adding a contribution 
to the diffusivity that is quadratic in $\phi$. In the present experiments, this effect is therefore expected to be negligible 
due to the smallness of $\phi$.

We have shown that the theoretical framework developed here and in our earlier paper (GC) is useful for analyzing 
real experimental data, not withstanding the fact that the theory pertains to very idealized situations. In particular, 
equation~(\ref{N}) is a simple generalization of the formula~(\ref{JLeq}) introduced by Jorgenson 
and Lukacs~\cite{jorgenson_zone_1981} that has proved to be extremely useful in laboratory practice. It is hoped 
that equation~(\ref{N}) could be used in a similar manner to characterize the effects of EMD.

\end{document}